\documentclass[oldversion]{aa}
\usepackage{natbib}
\usepackage{amssymb}
\usepackage{txfonts}
\usepackage{graphicx}

\newcommand{\Msun}{\ensuremath{\mathrm{M}_\odot}}
\newcommand{\Rsun}{\ensuremath{\mathrm{R}_\odot}}
\newcommand{\Lsun}{\ensuremath{\mathrm{L}_\odot}}

\newcommand\refereebf[1]{{#1}}
\newcommand\refereebfsecond[1]{{#1}}
\providecommand\beq{\begin{equation}}
\providecommand\eeq{\end{equation}}

\title{The evolution of runaway stellar collision products}

\author{
   Evert Glebbeek\inst{1}\thanks{Current address: Department of Physics and Astronomy, McMaster University, 1280 Main Street West, Hamilton, Ontario, Canada L8S 4M1} \and
   Evghenii Gaburov\inst{2,3}\thanks{Current address: Leiden Observatory, P.~O. Box 9513, 2300 RA Leiden, The Netherlands} \and
   Selma E. de Mink\inst{1} \and
   Onno R. Pols\inst{1} \and
   Simon Portegies Zwart\inst{2,3}
}

\institute{
Sterrekundig Instituut Utrecht, P.~O. Box 80000, 3508 TA Utrecht, The Netherlands.
\and
Sterrenkundig Instituut ``Anton Pannekoek'', Kruislaan 403, 1098 SJ Amsterdam, The Netherlands
\and
Section Computational Science, Kruislaan 403, 1098 SJ Amsterdam, The Netherlands
}

\abstract{
In the cores of young dense star clusters, repeated stellar collisions
involving the same object can occur. It has been suggested that this leads to
the formation of an intermediate-mass black hole.
To verify this scenario we compute the detailed evolution of the
merger remnant of three sequences, then follow the evolution until the onset
of carbon burning, and estimate the final remnant mass to determine the
ultimate fate of a runaway merger sequence.

We use a detailed stellar evolution code to follow the evolution of the
collision product. At each collision we mix the
two colliding stars, accounting for the mass loss during the collision.
During the stellar evolution we apply mass-loss rates from the literature,
as appropriate for the evolutionary stage of the merger remnant. We
computed models for high ($Z=0.02$) and low ($Z=0.001$) metallicity to
quantify metallicity effects.

We find that the merger remnant becomes a Wolf-Rayet star before the end of
core hydrogen burning. Mass loss from stellar winds dominates the mass
increase due to repeated mergers for all three merger sequences that we
consider. In none of our high-metallicity models an intermediate-mass
black hole is formed, 
instead our models have a mass of $10$--$14$ \Msun{} at the onset of carbon
burning.
\refereebf{For low metallicity the final remnant is more massive and may
explode as a pair-creation supernova.}
We find that our metal-rich models become inflated as a result of
developing an extended low-density envelope. This may increase the
probability of further collisions, but self-consistent $N$-body
calculations with detailed evolution of runaway mergers are required to
verify this.

}

\date{}

\bibpunct{(}{)}{;}{a}{}{,} 

\newlength{\timeswidth}
\newlength{\pluswidth}
\newcommand{\plustimes}{\ensuremath{%
\settowidth{\timeswidth}{$\times$}%
\settowidth{\pluswidth}{$+$}%
\addtolength{\timeswidth}{\pluswidth}%
+\hspace{-.5\timeswidth}\times%
}}

\begin{document}
\maketitle

\section{Introduction}\label{sec:introduction}

The usual mode of star formation leads to a spectrum of masses between
the theoretical hydrogen-burning limit and some upper limit, which
appears to be close to about
100\,\Msun\,\citep{1999ApJ...515..323E,2004MNRAS.348..187W,2005Natur.434..192F}.
In young and dense star clusters more massive stars can form when
two or more high-mass stars coalesce. The cluster environment helps in
driving these stars together. 
If this happens in a sufficiently young and dense star cluster the same star
may experience
multiple collisions in what is named a `collisional runaway'
\citep{1999A&A...348..117P}. During such a chain collision several
stars collide in short succession. The trigger for a chain collision
is the gravothermal collapse \citep{1984MNRAS.208..493B} of the core of
a young and dense star cluster. 
\refereebf{The density reached during core collapse depends on the cluster
parameters. For example, for the cluster MGG-11 in the starburst galaxy M82
\citep[$M = 3.5\times 10^5 \Msun$, $r = 1.2\,\mathrm{pc}$][]{2003ApJ...596..240M}, which was studied by \citet{2004Natur.428..724P}, this leads to a
central density $\rho_\mathrm{c} \simeq 10^8\,\Msun/\mathrm{pc}^3$.} If
cluster core collapse is initiated before the most massive stars leave the
main sequence \refereebf{(after about $3\,\mathrm{Myr}$,
\citealt{2004Natur.428..724P,2006ApJ...640L..39G})} the collisional runaway
sets in \citep{2002ApJ...576..899P} and continues until the target star
leaves the main sequence \citep{1999A&A...348..117P}.

If the star explodes as a supernova, this supernova is \refereebf{expected} to
be unusually bright and rich in hydrogen \citep{2007Natur.450..388P}.
The star may also collapse completely into a black hole, without a visible
supernova. The black hole remnant of such a star may be considerably more
massive than hitherto observed \citep{2004Natur.428..724P}, though less
massive than the supermassive black holes found in the nuclei of large
galaxies \citep{1995ARA&A..33..581K}. Various groups have now confirmed
the formation of such collisional runaways, and conjecture that the final
merger product collapses to a black hole of up to about $10^3$\,\Msun\,
\citep{1999A&A...348..117P,2004Natur.428..724P,2006MNRAS.368..141F,2006ApJ...640L..39G}.
\refereebf{Multiple collisional runaways within the same simulation were found by
\citet{2006ApJ...640L..39G} for simulations with a binary fraction larger than
0.1 and these authors argue that multiple collisional runaways may be common
in massive clusters.}

Two of our aims in understanding merger runaways are understanding the
structure of the merger remnant and the influence of stellar evolution of
the merger runaway.
Stellar evolution of very massive stars (with masses above $150\,\Msun$)
has recently been studied by \citet{belkus_evolmassivestars},
\citet{2008A&A...477..223Y} and \citet{langer_paircreationsn}.
These studies all seem to come to the same conclusion: at high metallicity
mass loss is copious enough to prevent the formation of a black hole of
more than $50\,\Msun$, which is much lower than the
conjectured intermediate-mass black hole mass.  
These studies, however, either used approximate formulae for stellar evolution
\citep{belkus_evolmassivestars} or studied the evolution of very massive
stars from the zero-age main sequence with an initially homogeneous
composition \citep{2008A&A...477..223Y,langer_paircreationsn}.
According to the dynamical simulations the massive star
grows in mass by means of repeated collisions with less massive
stars. The consequences of the collisions, the evolution between collisions
and the differences in stellar age and structure at the moment of
collision are not considered by these studies.
A first attempt to overcome these problems by calculating collisions
between massive stars and computing the evolution of these merged objects
did not result in very different conclusions \citep{suzuki_massivemerger}.
Ideally, one would like to perform a fully self-consistent simulation in
which the stellar dynamics, the hydrodynamics of the stellar collisions and
the further evolution of the collision products are taken into account self
consistently. Such multiscale simulations, however, will have to await the
development of the appropriate numerical methodology. The MUSE software
environment may provide the necessary functionality for such
simulations\footnote{see {\tt http://muse.li}.}.

In this work we investigate the evolution of three collisional runaways that occurred in
direct $N$-body simulations of the \refereebf{star cluster MGG-11
by \citet{2004Natur.428..724P}.  
From their simulations we extract the masses of the stars
involved in the collision sequences and the times of the collisions. 
We follow the evolution of the colliding stars with a stellar evolution
code (described in \S \ref{sec:evolutioncode}) until the moment of
collision.
The outcome of the merger event is modelled as described in \S
\ref{sec:collisions} and used as a new input model for the stellar
evolution code.
The evolution of the merged object is then followed until the next
collision.
Note that due to this approach the cluster dynamics is decoupled from the
evolution of the collisional runaway.}
After the last collision we follow the evolution of the merger product
until the evolution code fails to converge or until the onset of carbon
burning, applying mass loss rates from the literature. We estimate the
mass of the collision product at the end of the evolution and the mass of
the black hole remnant. We also provide the chemical yields that result from the
merger sequence and compare them to the combined yields of a population of
normal single stars. Finally we comment on the effect of the initial
composition, especially the heavy element content $Z$ (metallicity) on the
structure of the merger remnant.

\section{Methods}\label{sect:Methods}
\subsection{Stellar dynamics}
\refereebf{
The merger sequences used in this work were extracted from $N$-body
simulations of young star clusters using the {\tt kira} integrator of the
{\tt starlab} software environment \citep{2001MNRAS.321..199P}.
These simulations used 131072 stars and ran
on a special purpose GRAPE-6 computer \citep{2001ASPC..228...87M}.
In the original runs, stellar evolution was included using the prescription
of \citet{1989ApJ...347..998E}. Collisions were treated in a ``sticky
sphere'' approximation: when the distance between two stars becomes less
than the sum of their radii the stars are merged. The mass of the merged
object is the sum of the masses of the colliding stars. The evolution of
the merged star was continued using the same evolution prescription as used
for normal single stars.}

\subsection{Stellar collisions}\label{sec:collisions}
We use two different methods to model stellar collisions.
The first assumes that the collision product is in hydrostatic and thermal
equilibrium and mixed homogeneously.
The second method uses the prescription of \citet{2008MNRAS.383L...5G} to
model the structure of the remnant. In this case the collision product is
not homogeneously mixed and it is not in thermal equilibrium (although it
is still in hydrostatic equilibrium).

All collisions are treated as head-on collisions with vanishing velocity at
infinity (\emph{i.e.}, parabolic collisions). We ignore rotation in this
work despite the fact that rotation can have a significant influence on the
evolution of a massive star \citep{maeder_meynet_evolrotstars_review}.

\subsubsection*{Homogeneous mixing}
In this approach, detailed models of the progenitor stars were merged
and homogeneously mixed at each step of the sequence. We assume that the
merger remnant is in hydrostatic and thermal equilibrium. The mass loss from
the collision is estimated according to \citet{2008MNRAS.383L...5G}.

In general merger remnants are not fully mixed \citep{lombardi_collisions,
2008MNRAS.383L...5G, GlebbeekPols2008, GaburovGlebbeek2008}. However, 
\citet{suzuki_massivemerger} find that their massive collision products are
almost completely mixed during the subsequent evolution.
Similarly, we find (\S \ref{sec:structure}) that the central convection
zone in our merger remnants encompasses $\gtrsim 90\%$ of the
stellar mass.
Rotational mixing, which we have ignored, will result in even more extended
mixing of the collision product, so to first order homogeneous mixing is a
reasonable approximation for our models.

\subsubsection*{Detailed merger models}
Our detailed merger models were calculated using  Make Me A Massive Star
(MMAMS) \citep{2008MNRAS.383L...5G}.
The code has a prescription for the mass lost from the collision that is
based on the results of smooth particle hydrodynamics calculations. Heating
due to the dissipation of the kinetic energy of the progenitor stars in shocks
and tides is also taken into account.
After the ejected mass has been removed and heating has been applied to the
material from the parent stars the structure of the collision remnant is
determined by searching for a configuration that is dynamically stable
\citep[\emph{i.e.} satisfies the Ledoux stability
criterion,][]{book:kippenhahn_weigert}. An algorithm for doing this was
first developed by \citet{lombardi_mmas} for low-mass stars, for which it
is sufficient to sort the mass shells in order of increasing entropy and
then integrate the equation of hydrostatic equilibrium. 
For massive stars
where radiation pressure is important this does not necessarily produce a
stable configuration and some mass shells need to be moved again after
the equation of hydrostatic equilibrium has been integrated. This procedure is
repeated until a stable model is converged upon. 

The output model is imported into the stellar evolution code using the method
described in \citet{GlebbeekPols2008} and evolved until the time of the
next collision.

Due to heating during the collision the merger product is not in thermal
equilibrium. The excess of internal heat is radiated away
during the contraction of the star to the main sequence. Because the stars
were not in thermal equilibrium, we encountered more numerical problems
when importing the stellar models than we did for the homogeneously mixed
models.

\subsection{Stellar evolution}\label{sec:evolutioncode}
Our stellar evolution code is a version of the STARS code originally
developed by \citet{eggleton_evlowmass} and later updated
\citep[\emph{e.g.}][]{pols_approxphys}.
This version of the code uses the opacities from \citet{opal1996} that
take into account enhancement of C and O, as described in
\citet{eldridge_tout} and \citet{FergusonAlexander05}.
The assumed heavy-element composition is scaled to solar abundances 
\citep{anders_and_grevesse_abund}. 
Chemical mixing due to convection
\citep{bohm-vitense_convection,eggleton_mixingproc} 
is taken into account. 

STARS uses an adaptive mesh in which the mesh points automatically
redistribute themselves according to a mesh spacing function that places
more meshpoints in regions of the star where a higher resolution is
required. 
For the models presented here we used
$200$ mesh points for the main sequence phase and $500$ for the core
helium-burning evolution.
Because our stars form an extended low-density envelope (see \S
\ref{sec:structure}) we found it necessary to increase the number of
mesh points in the outer layers compared to our standard stellar models.

We use a mass fraction of heavy elements $Z=0.02$ for our standard runs. In
order to study the effect of metallicity we also recalculated one of
our sequences with $Z=0.001$. Metallicity affects the mass-loss rate and
therefore the mass of the progenitors at each collision.
We terminate the evolution at central carbon ignition and estimate the
final remnant mass according to \refereebf{the
prescription of} \citet{belczynski_remnantmasses}.

\subsection{Mass loss}
Since our collision products become very massive and luminous, even
exceeding the Humphreys-Davidson limit \citep[a
luminosity cutoff above which few stars are
observed,][]{humpreys_davidson}, mass loss plays a key role in their
evolution. Unfortunately, neither observations nor theoretical models of
mass loss exists for the full range of masses and luminosities reached by
our models. From the mass loss rates available in the literature we have
adopted the theoretical rate from \citet[][hereafter referred to as the
Vink rate]{vink_mass_loss,vink_mass_loss2} in the temperature range
$10\,000 \mathrm{K} \lesssim T_\mathrm{eff} \lesssim 50\,000 \mathrm{K}$.
For lower effective temperatures we use the empirical rate from
\citet[][hereafter referred to as the de Jager rate]{dejager_mass_loss}. 

All our models become helium-rich and evolve into Wolf-Rayet (WR) stars. We
follow the criterion used by \citet{2006A&A...452..295E} to decide when our
stars become WR stars. Specifically, we apply the
mass-loss rate from \citet{2000A&A...360..227N} when the surface abundance
of hydrogen drops below $0.4$ by mass \refereebfsecond{fraction} and
$T_\mathrm{eff} > 10\,000 \mathrm{K}$. In our models this happens
\emph{before} the star finishes core hydrogen burning 
\citep[as in][]{langer_paircreationsn,2008A&A...477..223Y}. For our low
metallicity run we used the metallicity scaling found by
\citet{vink_wrwindscaling} for the WR mass-loss rate. The
\citet{vink_mass_loss2} rate already includes metallicity scaling.

Apart from the de Jager and Vink rates, we also considered the mass
loss rate from \citet{kudritzki_mass_loss}.
None of these rates fully cover the range of parameters of our models. The
empirical de Jager rate has too few data points for luminosities above the
Humphreys-Davidson limit. Their 20 point Chebyshev fit needs to be
extended beyond $\log L/\Lsun = 6.7$. 
The Vink rate is based on atmosphere models calculated for $\log L/\Lsun
< 6.25$ and $10\,000 \mathrm{K} \lesssim T_\mathrm{eff} \lesssim 50\,000
\mathrm{K}$ and the Kudritzki rate needs to be extrapolated for
$T_\mathrm{eff} < 40\,000 \mathrm{K}$.

Our preference for the Vink rate comes from the
fact that it best covers the range of effective temperatures for our
models and can be better extrapolated in luminosity than the de Jager rate.
By contrast, the Kudritzki rate needs to be extrapolated to lower
effective temperatures, which is less reliable than extrapolating to higher
luminosities because the presence of spectral lines that drive the wind
is sensitive to the temperature. In the region where the
Kudritzki rate is valid it is very similar to our extrapolated Vink rate.
The Vink rate is not applicable to red supergiants and predicts a mass loss
rate that is too low for cool stars. For this reason we adopt the de Jager
rate rather than the Vink rate at effective temperatures below
$10\,000\mathrm{K}$. Our models only reach this temperature for
luminosities that are within the validity range of the de Jager rate.

Our adopted mass-loss rate likely underestimates the true mass loss
rate since our stars are much closer to their Eddington limit than the
model calculations on which the Vink rate is based. We will return to this
point in the discussion.

\subsection{Rotation}
It has been shown that for off-axis collisions the angular momentum of the
collision product can be so large that it cannot reach thermal equilibrium
before losing a large fraction of its angular momentum
\citep{lombardi_collisions,article:sills_evcolprod1}. We may therefore
underestimate mass loss from the collision.
The mechanism for this angular momentum loss is unclear but it has been
suggested that magnetic fields can play a key role
\citep{1995ApJ...447L.121L,2005MNRAS.358..716S}.
Rapid rotation can also enhance the mass-loss rate of a star, especially
close to the Eddington limit \citep{maeder_meynet_omega_limit}. This
increases the uncertainty in the mass-loss rate.

Rotation also influences the star through various instabilities that can
induce mixing \citep{1976ApJ...210..184E, pinsonneault_rotation,
heger_rotation}. This mixing is important because it can bring helium to
the surface, affecting the opacity of the envelope and increasing the
luminosity and effective temperature of the star.
As mentioned above, rotational mixing is not expected to alter the outcome
of our calculations very strongly because our collision products are almost
fully convective.

\begin{table}
\caption{
Parameters and results of the first collision sequence studied in this
paper. Time $t$ in Myr, masses are in solar units.
The final row gives the age at which our evolution calculations stopped and
the mass of the collision product at the end of the evolution (assumed
black hole mass).  
}
\label{tab:collision_table}
\begin{center}
\begin{tabular}{rrrrrrr}
\hline\hline
$N$ & $t$ & $M_{1,N\mathrm{body}}^\mathrm{a}$ &
$M_{1,\mathrm{mix}}^\mathrm{a}$ & $M_{1,\mathrm{MMS}}^\mathrm{a}$ &
$M_2^\mathrm{a}$ & \refereebf{$M_\mathrm{merger}^\mathrm{a}$} \\
\hline
  1 &  0.22577 &    92.4 &    90.9 &   90.9 &  79.4 &   154.3\\
  2 &  0.22782 &   171.8 &   154.3 &  152.8 &  85.3 &   217.1\\
  3 &  0.23702 &   257.1 &   216.7 &  213.6 &   7.7 &   223.3\\
  4 &  0.31217 &   264.8 &   219.8 &  215.3 &  77.8 &   274.4\\
  5 &  0.40122 &   342.6 &   268.2 &  261.6 &  13.8 &   279.7\\
  6 &  0.43480 &   356.4 &   277.2 &  268.0 &  71.9 &   327.1\\
  7 &  0.43480 &   428.3 &   327.1 &  315.1 &  79.4 &   382.2\\
  8 &  0.74123 &   507.7 &   348.2 &        &  30.1 &   371.8\\
  9 &  0.83549 &   537.7 &   360.2 &        &  66.7 &   406.7\\
 10 &  0.90575 &   604.5 &   397.3 &        &   1.2 &   398.5\\
 11 &  1.29919 &   605.7 &   353.5 &        &  98.2 &   415.2\\
 12 &  1.33450 &   703.9 &   411.0 &        &  24.2 &   430.8\\
 13 &  1.42468 &   728.0 &   417.6 &        &   8.9 &   425.7\\
 14 &  1.50261 &   736.9 &   409.0 &        &   2.2 &   411.2\\
 15 &  1.60190 &   739.1 &   216.7 &        &  60.2 &   256.5\\
 16 &  1.63186 &   799.4 &   221.0 &        &   9.2 &   228.9\\
 17 &  1.68112 &   808.5 &   186.0 &        &  81.1 &   237.1\\
 18 &  1.75684 &   889.6 &   177.3 &        &  40.9 &   205.5\\
 19 &  1.90688 &   930.5 &   193.6 &        &  31.6 &   216.3\\
 20 &  2.07245 &   961.9 &   129.9 &        &   2.6 &   132.2\\
 21 &  2.59778 &   963.1 &    47.6 &        &  55.2 &    88.6\\
 22 &  3.10786 &  1012.7 &    40.9 &        &  75.4$^\mathrm{b}$ &   110.1\\
    &  3.72835 &  1118.9 &    13.9 \\
\hline
\end{tabular}
\end{center}
\begin{list}{}{}
\item[$^\mathrm{a}$]
The primary mass according to the $N$-body code
$M_{1,N\mathrm{body}}$ and according to our fully mixed models
$M_{1,\mathrm{mix}}$, the mass of the secondary $M_2$ and the mass of
the fully mixed remnant after the collision $M_\mathrm{merger}$.
The mass $M_{1,\mathrm{MMS}}$ according to the MMAMS models we were able to
calculate is given for comparison.
\item[$^\mathrm{b}$] See text
\end{list}
\end{table}
\begin{table}
\caption{As Table \ref{tab:collision_table} for collision sequence 2}
\label{tab:collision_table2}
\begin{center}
\begin{tabular}{rrrrrr}
\hline\hline
$N$ & $t$ & $M_{1,N\mathrm{body}}$ & $M_{1,\mathrm{mix}}$ & $M_2$ & $M_\mathrm{merger}$\\
\hline
  1 &  0.72251 &    91.8 &    86.5 &  81.9 &   150.1\\
  2 &  0.72252 &   173.7 &   150.1 &  80.2 &   206.7\\
  3 &  0.88257 &   253.8 &   198.9 &  68.6 &   245.6\\
  4 &  0.88522 &   322.5 &   245.4 &  72.9 &   294.5\\
  5 &  0.95998 &   395.4 &   287.7 &   1.8 &   289.4\\
  6 &  1.41050 &   397.1 &   250.0 &  95.8 &   309.6\\
  7 &  1.50639 &   492.9 &   298.4 &  86.1 &   352.3\\
  8 &  1.97200 &   578.8 &   282.1 &  97.6 &   338.2\\
  9 &  2.22324 &   676.2 &    79.8 &  75.2 &   130.4\\
 10 &  2.22325 &   751.5 &   130.2 &  55.4 &   166.3\\
 11 &  2.45320 &   806.7 &    83.9 &  23.5 &   100.4\\
 12 &  2.68305 &   829.8 &    66.7 &  28.4 &    86.8\\
 13 &  2.74781 &   858.0 &    78.8 &   1.0 &    79.8\\
 14 &  3.08766 &   857.3 &    53.5 &  82.6 &   105.2\\
 15 &  3.19983 &   939.2 &    46.1 &  56.3 &    86.3\\
 16 &  3.27682 &   995.0 &    64.8 &  41.6 &    93.6\\
    &  4.51074 &  1036.6 &    10.1 \\
\hline
\end{tabular}
\end{center}
\end{table}
\begin{table}
\caption{As Table \ref{tab:collision_table} for collision sequence 3}
\label{tab:collision_table3}
\begin{center}
\begin{tabular}{rrrrrr}
\hline\hline
$N$ & $t$ & $M_{1,N\mathrm{body}}$ & $M_{1,\mathrm{mix}}$ & $M_2$ & $M_\mathrm{merger}$\\
\hline
  1 &  0.53120 &    77.8 &    75.3 &  70.7 &   131.5\\
  2 &  0.57294 &   148.5 &   130.7 &  82.8 &   191.3\\
  3 &  0.65612 &   231.3 &   188.1 &  58.5 &   228.6\\
  4 &  0.89154 &   289.8 &   215.4 &  96.6 &   279.9\\
  5 &  0.89155 &   386.4 &   279.9 &  78.5 &   332.5\\
  6 &  1.02208 &   464.9 &   318.1 &   1.5 &   319.5\\
  7 &  1.31086 &   466.4 &   289.3 &  56.3 &   327.9\\
  8 &  1.50575 &   522.6 &   305.7 &  48.9 &   340.1\\
  9 &  2.05327 &   571.3 &   164.0 &  72.7 &   209.2\\
 10 &  2.27966 &   643.8 &    73.1 &   2.0 &    74.8\\
 11 &  2.60213 &   645.0 &    35.1 &   3.5 &    37.7\\
 12 &  3.12235 &   642.8 &    19.8 &  49.3 &    60.7\\
    &  4.59400 &   692.1 &     9.7 \\

\hline
\end{tabular}
\end{center}
\end{table}
\begin{table}
\caption{As Table \ref{tab:collision_table3} but for $Z=0.001$.$^\mathrm{c}$}
\label{tab:collision_table3lowz}
\begin{center}
\begin{center}
\begin{tabular}{rrrrrr}
\hline\hline
$N$ & $t$ & $M_{1,N\mathrm{body}}$ & $M_{1,\mathrm{mix}}$ & $M_2$ & $M_\mathrm{merger}$\\
\hline
  1 &  0.53120 &    77.8 &    77.7 &  70.7 &   135.4\\
  2 &  0.57294 &   148.5 &   135.4 &  82.8 &   198.3\\
  3 &  0.65612 &   231.3 &   198.1 &  58.5 &   239.6\\
  4 &  0.89154 &   289.8 &   239.1 &  96.6 &   308.5\\
  5 &  0.89155 &   386.4 &   308.5 &  78.5 &   364.2\\
  6 &  1.02208 &   464.9 &   363.5 &   1.5 &   364.9\\
  7 &  1.31086 &   466.4 &   363.4 &  56.3 &   404.7\\
  8 &  1.50575 &   522.6 &   403.4 &  48.9 &   440.2\\
  9 &  2.05327 &   571.3 &   434.9 &  72.7 &   487.5\\
 10 &  2.27966 &   643.8 &   390.1 &   2.0 &   392.0\\
 11 &  2.60213 &   645.0 &   260.1 &   3.5 &   263.3\\
    &  2.83153 &   648.5 &   171.9$^\mathrm{d}$ \\
\hline
\end{tabular}
\end{center}
\end{center}
\begin{list}{}{}
\item[$^\mathrm{c}$] Note that the merger sequence was terminated earlier
than for $Z=0.02$ and that the merger remnant did not finish its evolution
before the evolution code broke down.
\item[$^\mathrm{d}$] Mass and time at the moment the evolution code broke
down, rather than the onset of carbon burning.
\end{list}
\end{table}
\begin{table}
\caption{Times $t$ and secondary masses $m_2$ for the sub merger sequence
leading to the secondary of collision 22 in the merger sequence in Table
\ref{tab:collision_table}.
}\label{tab:collision_tablex}
\begin{center}
\begin{tabular}{rrrrrr}
\hline\hline
$N$ & $t$ & $M_{1,N\mathrm{body}}$ & $M_{1,\mathrm{mix}}$ & $M_2$ & $M_\mathrm{merger}$\\
\hline
  1 &  0.88048 &   66.6  &   63.6  &  65.2 &  113.4 \\
  2 &  1.08887 &  131.8  &  110.4  &  16.4 &  114.1 \\
    &  3.10786 &  106.2  &   75.4  \\
\hline
\end{tabular}
\end{center}
\end{table}

\section{Results}\label{sec:results}

\begin{figure}
\includegraphics[width=0.5\textwidth]{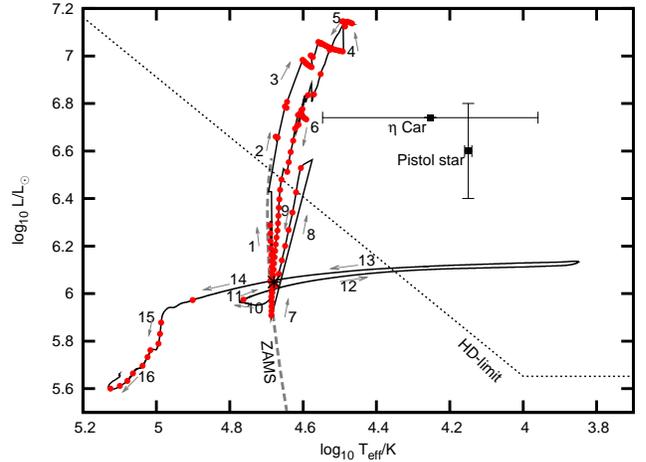}
\caption{
Evolutionary track of the merger from sequence 1 in the Hertzsprung-Russell
diagram.  The starting point is indicated by $\plustimes$ and
\refereebfsecond{numbered arrows indicate the way the star evolves along
the evolution track}. Points are
plotted on the evolutionary track after each $30\,000\,\mathrm{yr}$. The
dotted line indicates the Humphreys-Davidson limit. For reference, the
theoretical ZAMS (running up to 200 \Msun) and the locations of $\eta$
Carinae and the Pistol Star are also plotted.
}\label{fig:hrd2}
\end{figure}

\begin{figure}
\includegraphics[width=0.5\textwidth]{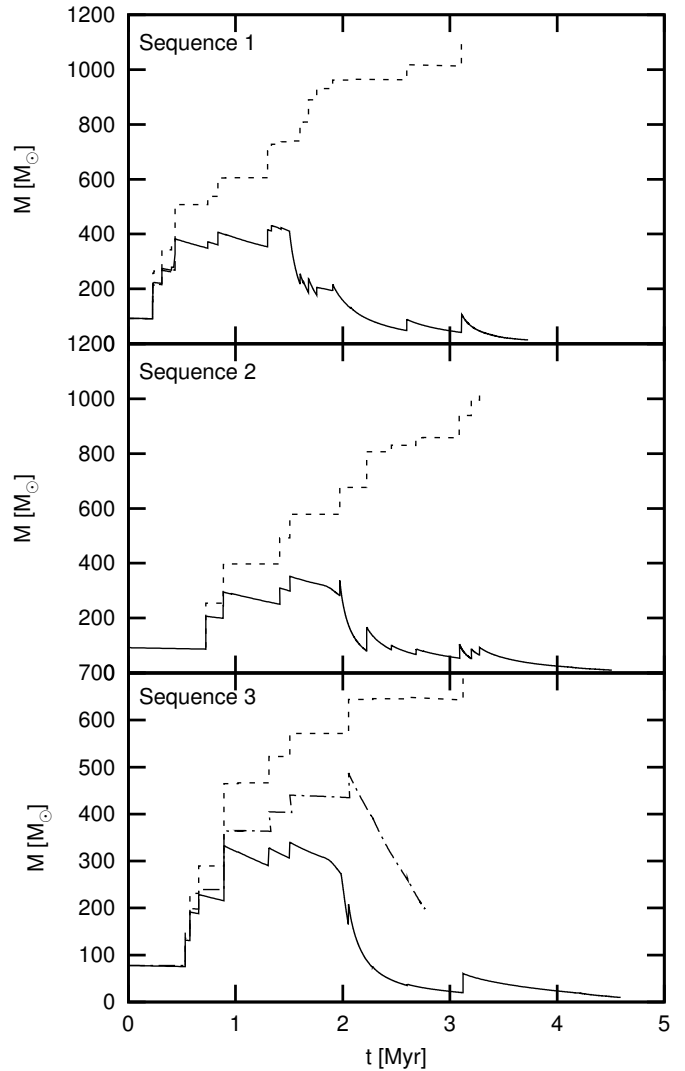}
\caption{The mass of the merger product as a function of time for the three
merger sequences listed in Tables
\ref{tab:collision_table}--\ref{tab:collision_table3}. The solid line is
the mass found from the detailed models assuming homogeneous mixing, the
dashed line is the mass predicted from the $N$-body calculation. The
dash-dotted line in the bottom panel is the mass of the $Z=0.001$ run.
}
\label{fig:mass_time}
\end{figure}

\begin{figure}
\includegraphics[width=0.5\textwidth]{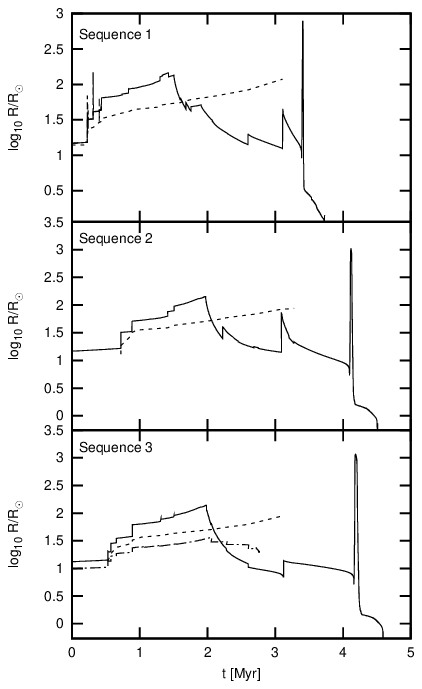}
\caption{
The radius of the merger remnant of the three sequences against time. The
solid line is the prediction from our homogeneous models while the dashed
line is the radius from the $N$-body code. The long dashed line in the top panel
is the radius from the MMAMS model, which shows spikes at each collision because
the merger remnant is out of thermal equilibrium immediately after the
merger. The dash-dotted line in the bottom panel is the radius of the
$Z=0.001$ run.
}\label{fig:radius_time}
\end{figure}

The initial conditions and outcome of each of our merger sequences are
listed in Tables \ref{tab:collision_table}, \ref{tab:collision_table2} and
\ref{tab:collision_table3}. The tables give the time of collision $t$
\refereebf{as well as the masses $M_1$ and $M_2$ of the colliding stars and
the mass $M_\mathrm{merger}$ of the merged object. In the following we
will refer to the object undergoing the subsequent merger events as the
primary and to the star colliding with it as the secondary. Note that the
mass of the primary depends on the treatment of the collisions and is
different for each run.} Sequence three was recalculated for $Z=0.001$ and
the results for this run are given in Table \ref{tab:collision_table3lowz}.

The evolutionary track in the Hertzsprung-Russell diagram (HRD) for the first merger
sequence (Table \ref{tab:collision_table}) is shown in Figure
\ref{fig:hrd2}. The two other sequences are similar.  The location of the
ZAMS (up to 200 \Msun) is indicated with a dashed line and every $30\,000
\mathrm{yr}$ is marked with $\bullet$. The repeated collisions drive the
collision product to high luminosities, exceeding the Humphreys-Davidson
limit, but the collision product never
moves far from the extension of the ZAMS, instead evolving nearly
vertically in the HR diagram \citep[similar to the evolutionary tracks for
homogeneously evolving stars,][]{yoon_chemical_homogeneous_evolution}.
For reference, the locations of the Pistol Star and $\eta$ Carinae are also
shown. The location of $\eta$ Car is based on \citet{hillier_etacar}, with
the errorbar in effective temperature due to the range in their radius
estimates. The luminosity is based on the infrared flux. The location of
the Pistol Star is based on the low luminosity solution of
\citet{figer_pistol_star}. We see that the merger remnant is always hotter
than either of these two stars, except when it becomes a red supergiant
(the red loop in Figure \ref{fig:hrd2}), at which time it is less luminous.
During the merger sequence the luminosity can exceed the luminosity of
these stars. On the other hand, $\eta$  Carinae and the Pistol Star are
both obscured by optically thick outflows, which means that comparing with
the effective temperature of our model can be misleading because we do not
model such an optically thick wind.  The location of the collision product
above the Humphreys-Davidson limit suggests that it is a luminous blue
variable (LBV) star, so that in reality its position in the
HRD is likely to be variable. 

The high luminosity increases the mass-loss rate, leading
to a competition between mass loss due to stellar winds and mass increase
due to collisions \citep[see also][]{suzuki_massivemerger,belkus_evolmassivestars}.
The time evolution of the mass of the mergers is shown in Figure
\ref{fig:mass_time}. The dashed lines give the mass that was predicted in
the $N$-body simulation \refereebf{(assuming no mass is lost in the
collision)} while the solid lines show our fully mixed models
for $Z=0.02$.
The two agree well for the first few collisions, but mass loss due to
stellar wind prevents the mass from exceeding $500\,\Msun$. The surface of
the merger remnants becomes helium-rich and after 1.5--2 Myr turns the star
into a Wolf-Rayet star. The strong WR mass-loss rate (up to
$3.6 \cdot 10^{-3}\,\Msun\,\mathrm{yr}^{-1}$ when the collision product first
becomes a WR star) brings the mass down on a timescale of $10^4$--$10^5
\,\mathrm{yr}$, to $\sim 100\,\Msun$
after the final collision and 10--14 \Msun{} at the time of carbon ignition.

In Figure \ref{fig:radius_time} we follow the evolution of the radius.
Despite the lower mass our
collision products have substantially larger radii (up to a factor three,
note the logarithmic scale) than was assumed by the $N$-body code.
This is due to a peculiarity in the structure of
the collision product, which will be discussed in detail in \S
\ref{sec:structure}. When the collision product becomes a WR star, the
radius decreases substantially and the collision product can be up to an
order of magnitude smaller than was assumed in the $N$-body calculation.
After core hydrogen exhaustion the collision product still has a thin
hydrogen-poor ($X \approx 0.04$) envelope. Expansion of this envelope is
responsible for turning the star into a red supergiant and causes the spike
in the radius at 3.5 Myr (first sequence) and 4.1 Myr (second and third
sequence). During the red supergiant phase the collision product is again
above the Humphreys-Davidson limit, but this phase is very short, lasting
$17 \cdot 10^{3} \mathrm{yr}$ ($<1\%$ of the lifetime of the star).

We have also plotted the mass and radius from the MMAMS models that we were
able to calculate. In the upper panel of Figure~\ref{fig:mass_time} the
MMAMS model is indistinguishable from the homogeneous model. After the collision
the merger remnant is out of thermal equilibrium and is inflated. The
increase in radius at each collision can be seen in the upper panel of
Figure \ref{fig:radius_time}. Once the collision product reaches thermal
equilibrium (after $\sim 10^4\,\mathrm{yr}$) the radius closely follows the
radius of the homogenised model, indicating that the homogenised model is
indeed a reasonable approximation of the structure of the merger remnant.
Because the radius is larger while the collision product is out of thermal
equilibrium it is more likely to interact or collide with other stars at
this time, but since we do not take feedback on the dynamics into account
this effect is not important for our present considerations.

The collision sequences mostly involve main-sequence stars but a few of the
listed collisions are special.
Collision 22 of the first sequence involves the remnants of two collision
runaways. The star that has undergone the longest sequence of collisions
(the ``primary'') is an early type Wolf-Rayet star at this point with a surface
hydrogen abundance of $0.24$ and a mass of $41\,\Msun$. However, it is
still undergoing core hydrogen burning. The star that has
undergone the shortest merger sequence (the ``secondary'', see Table
\ref{tab:collision_tablex}) is a core helium
burning star of $75\,\Msun$ that has not yet become a Wolf-Rayet star
although its surface hydrogen abundance is $0.44$, which is close to our
criterion.
In our homogeneous mixing treatment the
result is a collision product that has been enhanced in carbon (see the
surface abundance plot in Figure \ref{fig:surface_composition}), which is
converted into nitrogen through CNO processing.
In a more detailed treatment of the merger process we expect the dense
helium core of the secondary to sink to the centre of the collision
product so that the merger remnant would have a hydrogen depleted core.

A similar situation occurs for collision 14 from the second sequence, for
which the secondary has also become a core helium-burning star at the time
of collision.

\subsection{Structure and size of the merger remnants}\label{sec:structure}
\begin{figure}
\includegraphics[width=0.5\textwidth]{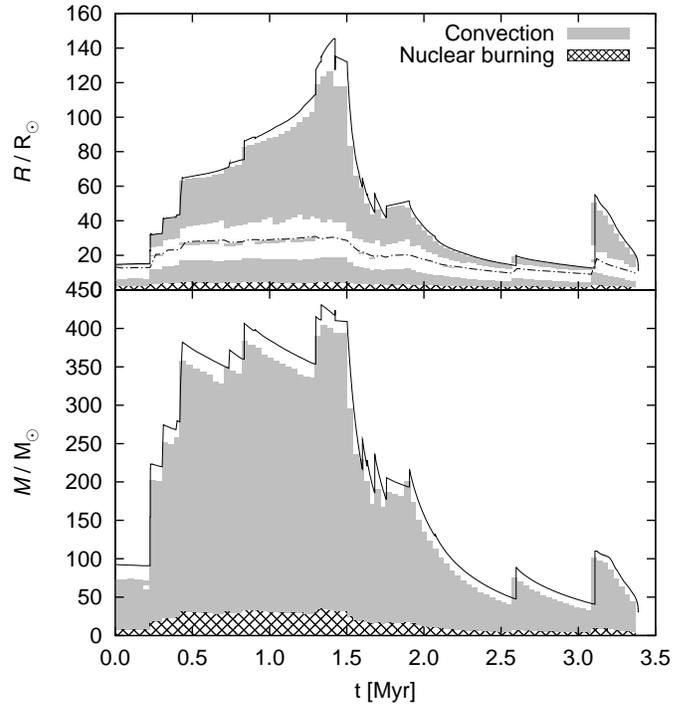}
\caption{
Kippenhahn diagram showing the evolution of the collision product of
the first merger sequence. The plot shows convection zones against
the radius (upper panel) and against mass coordinate (lower panel) as
a function of time. In the upper panel the radius containing a fraction
\refereebf{0.99999} of the stellar mass is indicated with a dash-dotted
line. The convective core encompasses 90\% of the mass but a much smaller
fraction of the radius (at most $20\,\Rsun$). Mass loss between collisions
exposes material from the convective core.
}\label{fig:kippenhahn}
\end{figure}

Apart from the mass the size of the merger remnant is one of the parameters
that determines the probability of subsequent collisions in a cluster. As
mentioned, the large radius of the $Z=0.02$ models during the first 1.5--2
Myr in Figure \ref{fig:radius_time} is caused by a peculiarity of the
stellar structure.

The merger remnants become very massive and are almost fully convective.
The lower panel in Figure \ref{fig:kippenhahn} shows the location
of convection zones against the enclosed mass, the upper panel shows the
same information as a function of the radius.  The central convection zone
contains about 90\% by mass of the merger remnant.
As can be seen from the figure, mass loss between the
collisions can reveal material from the central convection zone at the
stellar surface (for instance, the mass of the star at 1.3 Myr is lower
than the mass of the convective core at 0.8 Myr).

The outer 10\% by mass of the merger remnant is radiative, but the
upper panel of Figure \ref{fig:kippenhahn} shows two convection zones in
this region which correspond to different peaks in the opacity of the
stellar material.
An extended convective layer that corresponds to the ``Fe bump'' is located
at large radii. The Fe bump is an increase in opacity around $\log T
\approx 5.3$ which was found after introducing the treatment of spin-orbit
splitting of iron and nickel into the computation of the opacity tables
\citep{RogersIglesias1992, Seaton1994}.  Deeper down a thin convective layer can
be seen, caused by the ``deep Fe bump'', occurring around $\log T \approx
6.3$.

Together, these convection zones are very extended in radius but contain
almost no mass.
It is especially the convection zone corresponding to the
Fe-bump which expands even more while the star evolves. When the star reaches
its maximum radius of about 150 \Rsun{} after 1.4 Myr, this convection zone
extends over 90 \Rsun{}, while it contains only about $10^{-4}\,\Msun$. At
this moment the star consists of a core of less than 30 \Rsun{} in size
containing almost all of the mass surrounded by an extended ``halo''
reaching from 30 to 150 \Rsun{}. This halo has an almost constant
temperature and density of $10^{-10}\,\mathrm{g\,cm^{-3}}$.
This is indicated in Figure \ref{fig:kippenhahn}, which shows the radius
outside which a fraction $10^{-5}$ of the stellar mass is located.

This ``core-halo'' structure has been found before in models of massive
stars, for example by \citet{Ishii_1999} for hydrogen-rich stars and
by \citet{petrovic_inflated_wr} for helium stars.
\citet{petrovic_inflated_wr} note that to provide the high mass-loss rate
from the surface a high outward velocity is needed in the outer layers
where the density is low. In their models the necessary velocity is higher
than the local sound speed by an order of magnitude. This means that the
halo cannot be modelled realistically under the assumption of hydrostatic
equilibrium and may not be stable. They find that with a more detailed
treatment the halo disappears as a result of the high mass-loss rate.
Because our merger remnants have a much larger radius the outflow velocity
in the halo is about 2--3 orders of magnitude smaller than
the sound speed, which suggests that the halo structure is stable in this
case.
The halo disappears when the merger remnants become Wolf-Rayet stars and
the mass-loss rate increases.

\subsection{Final remnant masses}
For each of the three merger sequences the collision product is close to
core hydrogen exhaustion when the merger sequence ends.
After the end of the main sequence hydrogen continues to burn in a shell very
close to the surface. The hydrogen envelope expands, driving the star into
a red loop in the HRD. Mass loss from the surface gradually removes the hydrogen
envelope, reducing the efficiency of the hydrogen burning shell. When the
hydrogen shell is extinguished the star returns to the blue part of the
HRD. The remaining hydrogen envelope is lost and the
star becomes a massive ($\sim 20-30\,\Msun$) helium star.
By the end of core helium-burning the mass has gone down to $10-14\,\Msun$,
$80\%$ of which is taken up by the C/O core. The expected outcome of the
evolution for such stars is a complete collapse to a black hole.
In each of these sequences $600$ -- $900\,\Msun$ is
lost to the interstellar medium (see Table \ref{tab:yields}).

\begin{table*}
\caption{Ejected mass and composition for the three computed merger
sequences compared to \refereebfsecond{the winds of} a population of
single stars \refereebfsecond{corresponding to sequence 1}.
}
\label{tab:yields}
\begin{tabular}{r|r|rrrr|rrrr|rrrr}
\hline
 & Single & \multicolumn{4}{c}{Sequence 1} & \multicolumn{4}{|c}{Sequence
 2} &\multicolumn{4}{|c}{Sequence 3} \\
 & \refereebfsecond{Wind} & Coll. & Wind & Remnant & Total & Coll. & Wind & Remnant & Total & Coll. & Wind & Remnant & Total \\
\hline\hline
$\Delta M$   &  647.5 &  228.6 &  695.8 &  95.4  &   1020 &  191.8 &  549.9 &   83.5 & 825   &  142.3 &  456.6 &  102.0 &  701  \\
H            & 0.4806 & 0.5648 & 0.4006 & 0.0899 & 0.4083 & 0.5599 & 0.3489 & 0.1944 &0.3823 & 0.5521 & 0.3361 & 0.2025 & 0.3605\\
He           & 0.4965 & 0.4000 & 0.5794 & 0.8308 & 0.5627 & 0.4093 & 0.6311 & 0.7637 &0.5930 & 0.4305 & 0.6439 & 0.7553 & 0.6168\\
C            & 0.0040 & 0.0012 & 0.0003 & 0.0306 & 0.0033 & 0.0090 & 0.0003 & 0.0133 &0.0036 & 0.0004 & 0.0003 & 0.0184 & 0.0030\\
N            & 0.0092 & 0.0166 & 0.0132 & 0.0294 & 0.0155 & 0.0117 & 0.0133 & 0.0128 &0.0129 & 0.0105 & 0.0134 & 0.0123 & 0.0127\\
O            & 0.0040 & 0.0013 & 0.0010 & 0.0060 & 0.0015 & 0.0044 & 0.0009 & 0.0081 &0.0024 & 0.0013 & 0.0008 & 0.0050 & 0.0015\\
\hline
\end{tabular}
The first row
lists the total mass $\Delta M$ (in solar units) lost through each of the
three listed channels, the remaining rows give the abundances (by mass
\refereebfsecond{fraction}) of H, He, C, N and O. For each sequence the
first column lists the ejecta from the collision, the second column lists
the integrated values for the stellar wind during the merger sequence,
the third column lists the values for the evolution of the merger remnant
after the merger sequence ended and the fourth column gives the
combined yields from these channels.
\end{table*}

\subsection{Surface abundances and chemical yields}\label{sec:yields}
\begin{figure}
\includegraphics[width=0.5\textwidth]{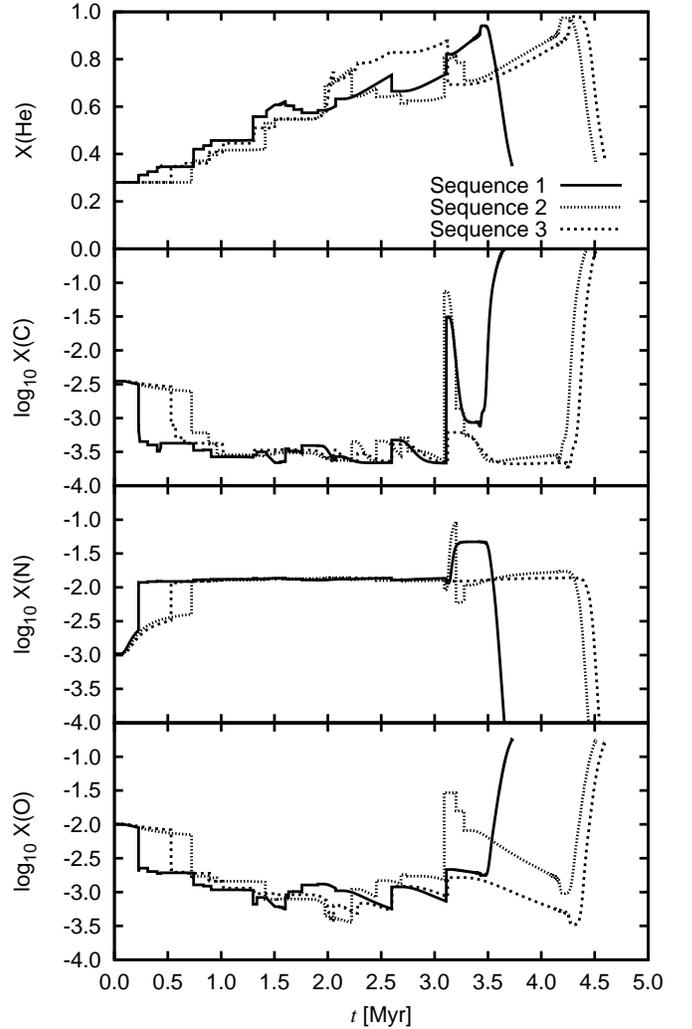}
\caption{
Surface He, C, N and O abundances (by mass fraction) as a function of
time (in Myr) for the three $Z=0.02$ sequences.
}\label{fig:surface_composition}
\end{figure}
In the course of its evolution the surface of the merger remnant gradually
becomes helium-rich, as can be seen in the top panel of Figure
\ref{fig:surface_composition}. The CNO abundances (shown in the bottom
panels of Figure \ref{fig:surface_composition}) change strongly at the
first collision and then stay mostly constant up to $t\approx
3\,\mathrm{Myr}$.

The abundances change most strongly during collisions. This is
because the merger remnant is fully mixed at this stage. The abundances
also change in between the collisions, as mass loss strips away the surface
layers and reveals the deeper layers, but the change is not visible on the
scale of the plots until the merger remnant becomes a Wolf-Rayet star. In
sequences 1 and 2 the merger remnant undergoes a collision with a core
helium-burning star at $3 \mathrm{Myr}$. This results in a strong increase
in the carbon abundance (and oxygen, for sequence 2) and a decrease in the
nitrogen abundance (through dilution). CNO cycling then converts the carbon
into nitrogen, producing a nitrogen-rich WR star.
However, our assumption of complete mixing is unlikely to be valid for
collisions with core helium-burning stars. More likely, most of the carbon
and oxygen would remain in the core of the collision product and such a
strong increase in surface C and N probably does not occur.

As the collision product continues to evolve after the end of the merger
sequence, the surface helium abundance increases until the hydrogen envelope
has been lost and the surface is nearly pure helium. The stellar wind
continues to expose deeper layers of the star, eventually revealing at the
surface the ashes of helium burning. At this point, the surface nitrogen
and helium abundance decrease while the carbon and oxygen abundances
increase strongly. At the end of the evolution, carbon is the most abundant
element on the surface.

The expulsion of gas from the cluster is usually attributed to supernova
explosions, which are expected to start after about $3\,\mathrm{Myr}$. The
merger remnant loses most of its mass before this time. 
Table \ref{tab:yields} gives the composition of
the material lost from the merger remnant as well as the amount of material
lost, split into three categories: ejecta resulting from the collisions,
mass loss due to stellar wind during the merger sequence and mass loss
\refereebf{due to stellar wind during the remaining evolution after the last
collision.} Most of the material is ejected in the form of a stellar wind
between collisions, followed by the material that is lost during the
collisions.
The material that is lost by the collision product after the end of the
merger sequence is significantly more helium-rich than the material that
was lost before, which simply reflects the increased surface helium
abundance of the collision product. The material lost from the collisions is
less helium-rich than the material that is lost in the wind. This is
partially due to the increase in the surface helium abundance between
collisions (Figure \ref{fig:surface_composition}) and partially due to the
fact that the estimated mass loss from the collision is larger for more equal
masses and becomes smaller when the mass ratio is more extreme, which is
the case for later collisions when the collision product is both massive
and helium-rich.

\refereebf{In column 2 of Table \ref{tab:yields} we give the total yield
that would have been obtained from stellar winds if we had followed the
evolution of the stars in the collision sequence individually. Because we
only follow the evolution until the onset of carbon burning we only compare
the pre-supernova yields.}
We first note that
the single stars eject much less material than the merger sequence. This is
because the merger sequence produces one $13.9\,\Msun$ black hole, while
the single stars above $30\,\Msun$ all produce black holes of 8--24
\Msun.
The ejected material is also less helium-rich than the material that is
lost from the merger remnant: the single star models lose 321 \Msun{} of 
helium and 311 \Msun{} of hydrogen ($\mathrm{He}/\mathrm{H} \approx 1$),
while the merger product loses 574 \Msun{} of helium and 416 \Msun{} of
hydrogen ($\mathrm{He}/\mathrm{H} \approx 1.4$). This is directly
related to the large size of the convective core: for the merger
remnant, $90\%$ of the material has undergone nuclear processing in the
core, which is a much larger fraction than for the population of single
stars.

\refereebfsecond{To compare the total yields from the merger sequence and
the single stars we should, of course, also include the supernova yields.
Here we can unfortunately only be qualitative because we do not have 
supernova yields for the merger remnant from our calculations and cannot
rely on models from the literature. Supernova yields of normal massive single
stars are of course available from the literature
\citep[\emph{e.g.}][]{1995ApJS..101..181W, 2004ApJ...608..405C,
2005A&A...433.1013H}.}

\refereebfsecond{Qualitatively, the net effect of the merger sequence is to
suppress supernova nucleosynthesis: whereas the single stars produce
multiple supernovae, the merger sequence produces only one supernova. In
particular the merger sequence will produce less oxygen than a population
of single stars.}
\subsection{Metallicity effects}\label{sec:metallicity}
\begin{figure}
\includegraphics[width=0.5\textwidth]{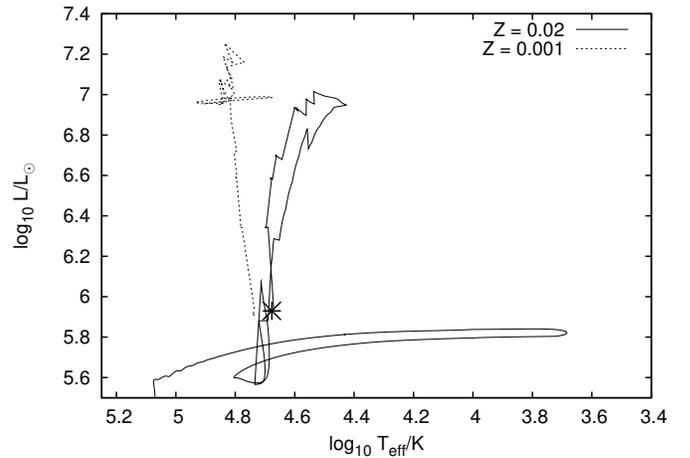}
\caption{
Evolutionary track of the merger remnant of sequence 3 for $Z=0.02$ (solid
line) and $Z=0.001$ (dashed line). Note that the $Z=0.001$ track remains
much brighter than the $Z=0.02$ track.
}\label{fig:hrdz}
\end{figure}

Because mass-loss rates are lower at low metallicity we recalculated
sequence 3 for $Z=0.001$. At this metallicity the mass-loss rates are about
13 times lower than for $Z=0.02$, which means that the remnant can become
more massive.

The mass of the remnant after the last collision in the sequence is $263
\Msun$. 
We followed the evolution of the merger product through core helium burning
until the core helium abundance decreased to $Y_\mathrm{c} \approx 0.63$
(about one third of the core helium-burning lifetime), at which time the
mass was $172\,\Msun$.  Numerical problems prevented us from evolving this
model further until carbon ignition.
We extrapolated the mass loss as a linear function of the central helium
abundance (rather than time) to the moment of helium exhaustion, which was
found to give a good estimate of the final mass for $Z=0.02$.  Based on
this extrapolation we expect the final mass to be $\sim 120\Msun$ by the
time of carbon ignition. The expected fate of this merger remnant is a
pair-creation supernova \citep{langer_paircreationsn, 2007Natur.450..388P}.

Because the opacity bump that gives rise to the core-halo structure is
associated with iron, the core-halo structure does not appear at low
metallicity and the collision product remains more compact. This is why in
the bottom panel of Figure \ref{fig:radius_time} the radius of the
$Z=0.001$ model (dash-dotted line) is smaller than the radius of the
$Z=0.02$ model until 2 Myr, when the latter becomes a WR star. The radius
is always smaller than the radius assumed in the $N$-body calculations.

The lower mass-loss rate makes it easier to build up more massive remnants,
but the smaller radius reduces the probability of a collision and may
prevent the occurrence of a runaway merger. The thermal timescale is also
shorter at low metallicity, meaning that the star will reach thermal
equilibrium faster after the collision.  An additional complication is that
the lifetime of the stars is reduced at lower metallicity (because of their
higher mass) so that there is less time to form the merger sequence.  Fully
self-consistent dynamical models (in which the stellar evolution, stellar
collisions and stellar dynamics are all treated consistently) are necessary
to determine which of these effects dominates in practice.

\section{Discussion and conclusions}\label{sect:Discussion}

\subsection{Mass loss and rotation}
The main uncertainty in our models is the adopted mass-loss rate. We have
used established mass-loss prescriptions from the literature.
Our adopted \citet{vink_mass_loss2} rate is derived for stars that are further
away from their Eddington luminosity than our collision products and is
likely to be an underestimate of the true mass-loss rate for our stars.

We apply WR mass loss rates when the surface hydrogen abundance
$X_\mathrm{s} < 0.4$ .
This transition is somewhat arbitrary and ideally we would use
a single mass loss recipe that predicts the mass-loss rate as a function of
the local luminosity, temperature, effective gravity and composition. No
such recipe is available in the literature at the time of writing.

The mass-loss rate can become very high during the WR phase due to our
extrapolation of the empirical rate to higher luminosity than for
which it was derived. We made sure that the power used to drive
the wind is always less than that provided by the star's luminosity
\citep{2004ApJ...616..525O}.
When the collision products first become WR stars, the mass-loss rate is
still very high. In part this is due to the sudden transition to WR mass
loss rates when $X_\mathrm{s}$ drops below 0.4. We expect that if the
transition is made more smoothly the mass-loss rate would increase earlier
and avoid the high peak value found in our current models. This again
stresses the need for a unified single mass loss prescription.

Our collision products have luminosities close to their Eddington
luminosity and for a substantial amount of time exceed the
Humphreys-Davidson limit.
Stars close to this limit become luminous blue variables which
can lose a large amount of mass in outbursts.  
A model that describes mass loss from stars that exceed their Eddington
luminosity is the so-called porosity model \citep{2004ApJ...616..525O,
2008arXiv0801.2519O}. Our models come close, but do not exceed their
Eddington luminosity. We made one trial run with the mass-loss rate
artificially enhanced by a factor of $50$ and found that 
\refereebf{although the star acquires new material from collisions this is
not enough to overcome the high mass loss rate and the star is very quickly
stripped of its mass}.

We have ignored rotation in this work. Rotational mixing is expected to be
unimportant because our models are almost fully convective. Rapid rotation
is also expected to enhance the mass-loss rate from stars, especially close
to the Eddington limit \refereebf{\citep{maeder_meynet_omega_limit}}.
\refereebf{Off-axis collisions almost always produce remnants that
need to lose angular momentum before they can reach thermal equilibrium
\citep{2005MNRAS.358..716S}. How this angular momentum is lost and how much
mass is lost in the process is currently poorly understood, further
increasing the uncertainty in the mass loss rate.}

\subsection{Implications for nucleosynthesis}
\refereebf{Anomalous chemical abundances in globular clusters are the topic
of much debate in the literature. Some stars show N and Na enhancement but
O depletion and some show indications of He enhancement. Current scenarios
that attempt to explain these chemical anomalies require pollution of the
interstellar medium by massive stars \citep{2007A&A...475..859D} or AGB
stars \citep[\emph{e.g.}][]{2008arXiv0809.1438D} in the central regions of
the cluster and a second episode of star formation, followed by the loss of
most of the first generation of low mass stars. In this context it is
interesting to consider the abundance pattern that would result
from a collisional runaway, not just to compare it with other scenarios but
also because this can give us a chemical signature that would allow us to
determine whether or not a runaway occurred in the past in a cluster
observed today.

The simplified nuclear network in our evolution code only
follows the main isotopes of the elements listed in Table \ref{tab:yields}.
In particular, we do not follow the evolution of isotopes of Na, Mg or Al.
However, we can make a few qualitative estimates.
The material that is lost from the collision remnant is helium-rich and
shows the signature of CNO processing (N enhancement, C depletion). The
material is also ejected in the centre of the cluster. In this sense, the
expected abundance pattern would be similar to that generated by the other
scenarios mentioned above. We do expect the collision sequence to produce
less oxygen and metals than a population of single stars because in this
case there is only one star that can generate a supernova and that will not
eject a large amount of material, given that its pre-supernova mass is
very small.
Nevertheless it is difficult to identify how it could be distinguished from
the other scenarios observationally without more detailed nucleosynthesis. 
It is possible to infer yields for a wider range of isotopes from our models
using a nucleosynthesis post-processing code. We plan to present the
results of such a study, \refereebfsecond{including an estimate for the
supernova yields} in a follow-up paper.

Although the collisional runaway causes more
material to be returned to the interstellar medium than the equivalent
population of single stars would have done, the amount of ejected material is
still small: observed helium-rich populations generally seem to
comprise about 15--20\% of the stars in the cluster
\citep{2008ApJ...672L..25P}, which is larger than the fraction of mass
lost from the merger remnant compared to the total cluster mass. Smilar to
other scenarios, this scenario requires that many first generation stars
are lost from the cluster. It appears, however, that this may be feasible
\citep{2008arXiv0809.1438D}. Multiple collisional runaways, as found by
\citet{2006ApJ...640L..39G}, can of course increase our estimate of the
yields.
}

\subsection{Conclusions and outlook}

We find that the end result of a runaway merger at solar metallicity
is a $\sim 100\,\Msun$ WR star after the final collision that produces a
$\sim 10\,\Msun$ black hole. Most of the mass is lost in the form of a
stellar wind enriched in N and He. At lower metallicity the mass-loss rates
are reduced and the remnant mass can be higher ($\sim 260\,\Msun$),
possibly leading to a pair-creation supernova. 
\refereebf{In all cases strong mass loss, in particular during the WR
phase, prevents the formation of an intermediate-mass black hole. Because
we have decoupled the $N$-body dynamics from the stellar evolution in this
work the mass and radius of our detailed evolution models are different
from those used in the $N$-body run, which would change the collision
probability.
We do not expect this to alter our conclusions because the physical
mechanism that prevents the formation of the intermediate-mass black hole,
the strong WR wind, does not depend on the dynamics.}

Nevertheless it is important to perform fully self-consistent simulations of
collisional runaways.
Calculations like those by \citet{belkus_evolmassivestars}  in which the
stellar evolution is treated with an analytic recipe, are an important step
in this direction. We will extend this in future work by coupling our
stellar evolution code to a dynamics code in the MUSE framework
\citep{muse_paper}.
Calculations at lower $Z$ are especially interesting since remnant masses
can be higher due to a reduced stellar wind, but on the other hand
collisions are less likely because the stars are more compact.

\section*{Acknowledgements}
Evert thanks Jorick Vink and Allard Jan van Marle for useful discussion
about mass-loss rates from luminous stars in general and Henny Lamers for
the Wolf-Rayet mass-loss rates in particular. He also thanks Rob Izzard for
more general discussion. \refereebf{We thank Georges Meynet for providing
many useful and inspiring comments that greatly improved this paper.}

EG\&EG are supported by NWO under grants 614.000.303 and 635.000.303. SdM
is partially supported by NOVA.  The research was conducted in the context
of the MODEST collaboration.

\bibliographystyle{aa}
\bibliography{GGPPZ2007l}

\end{document}